\documentclass[a4paper]{article}

\usepackage{CJK,multicol,multirow,graphics}
\usepackage[a4paper,margin=2.5cm]{geometry}
\usepackage{xcolor}
\usepackage{amsmath,amsfonts,amssymb,bm,mathrsfs}
\usepackage{graphicx}
\usepackage{booktabs}
\usepackage[font=footnotesize,labelfont=bf,labelsep=period]{caption}
\usepackage[colorlinks=true,linkcolor=blue,citecolor=blue,urlcolor=blue]{hyperref}
\usepackage[style=phys,sorting=none,giveninits=true,articletitle=false,maxbibnames=10,doi=false,url=false,isbn=false]{biblatex}

\graphicspath{{figures/}}
\renewcommand{\cite}{\supercite}

\newcommand{\manuscripttitle}{Colossal Spin--Lattice--Charge Coupling in Noncollinear Ferrimagnetic NbMnX$_2$ ($X$ = S, Se, Te) Monolayers}

\hypersetup{
  pdftitle={Colossal Spin-Lattice-Charge Coupling in Noncollinear Ferrimagnetic NbMnX2 (X = S, Se, Te) Monolayers},
  pdfauthor={Peng Wang, Nanshu Liu, Hongkuan Yuan, Xue Jiang, Jijun Zhao}
}

\begin{document}

\begin{center}
\vspace*{-4mm}
{\large\bfseries\boldmath\manuscripttitle\par}
\footnotetext{$^{*}$Corresponding authors. Emails: \href{mailto:jiangx@scnu.edu.cn}{jiangx@scnu.edu.cn}; \href{mailto:zhaojj@scnu.edu.cn}{zhaojj@scnu.edu.cn}.}
\par\vspace{5mm}
{\normalsize\rm Peng Wang$^{1}$, Nanshu Liu$^{1}$, Hongkuan Yuan$^{1}$, Xue Jiang$^{2,3*}$, and Jijun Zhao$^{2,3*}$\par}
\par\vspace{3mm}
{\small\slshape $^{1}$Chongqing Key Laboratory of Micro \& Nano Structure Optoelectronics, School of Physical Science and Technology, Southwest University, Chongqing 400715, China\par}
{\small\slshape $^{2}$Guangdong Basic Research Center of Excellence for Structure and Fundamental Interactions of Matter, Guangdong Provincial Key Laboratory of Quantum Engineering and Quantum Materials, School of Physics, South China Normal University, Guangzhou 510006, China\par}
{\small\slshape $^{3}$Guangdong-Hong Kong Joint Laboratory of Quantum Matter, Frontier Research Institute for Physics, South China Normal University, Guangzhou 510006, China\par}
\end{center}

\small{\narrower
Noncollinear ferrimagnets combine a net magnetic moment with internal spin degrees of freedom, creating opportunities to couple magnetic, mechanical, thermal, and electronic responses. Using first-principles calculations, symmetry analysis, atomistic spin simulations, and transport modeling, we identify stable NbMnX$_2$ ($X$ = S, Se, Te) monolayers with noncollinear ferrimagnetic ground states and Curie temperatures up to 313 K. Competing exchange, Dzyaloshinskii--Moriya interactions, and magnetic anisotropy produce field-responsive canted states that couple strongly to the lattice and charge carriers. Magnetization rotation changes the elastic modulus of NbMnTe$_2$ by up to 36\%, while lattice entropy substantially enhances the near-room-temperature magnetocaloric response of NbMnS$_2$ and NbMnSe$_2$. In NbMnSe$_2$, rotation restores a mirror-protected crossing, continuously closes an approximately 0.13 eV gap, and produces a conductivity-based anisotropic magnetoresistance ratio up to 10,000\% at 100 K. The combined results establish noncollinear NbMnX$_2$ monolayers as a multifunctional platform for two-dimensional spintronics, nanomechanics, and thermal control.
\par}
\vskip 3mm
\normalsize\noindent{\narrower{\textit{Keywords}: noncollinear ferrimagnetism; two-dimensional magnets; magnetoelastic coupling; metal--insulator transition; magnetocaloric effect; anisotropic magnetoresistance\par}}

\section*{Introduction}

The discovery of intrinsic long-range magnetic order in two-dimensional (2D) materials has opened a route to devices in which magnetic states can be manipulated at the atomic scale.\cite{gong20191,jiang20212} Most known 2D magnets are collinear ferromagnets, antiferromagnets,\cite{olsen20243,mu20264} or altermagnets.\cite{song20255,bai20246} Noncollinear ferrimagnets (NC-FiMs) are much less common, yet they combine a finite net moment with reduced symmetry and competing interactions.\cite{li20237,song20228,kim20269,andriushin202510,zhang202511,gonzlezhernndez202412} Their internal spin coordinates can therefore respond to magnetic fields while coupling strongly to lattice and electronic degrees of freedom.\cite{telford202213,jia202514}

Such coupling underlies magnetoelasticity,\cite{cheng202617,chen202518,berk201919} magnetocaloric effects (MCEs),\cite{patra202320,martinhovieira202221} anisotropic magnetoresistance (AMR),\cite{ritzinger202322} and magnetization-driven electronic phase transitions. Realizing these responses together in one near-room-temperature 2D magnet would enable magnetic control of mechanics, heat, and charge, but no material platform combining NC-FiM order, large magnetoelastic and magnetocaloric responses, and a magnetization-driven metal--insulator transition (MIT) has been established.

Layered double transition-metal chalcogenides M$'$M$''$X$_2$ provide a promising setting because their inequivalent transition-metal sublattices combine localized 3$d$ moments with more extended 4$d$/5$d$ states.\cite{wang201723,bai201924,li201925,neu201926,mazzola202327,mei202328,stepanova202429,wu202430,gao202531,li202632,ershadrad202633,jiao202634,su202607} Competing exchange and spin-dependent hybridization can coexist with strong spin--orbit coupling (SOC) and symmetry-allowed relativistic interactions. Related compounds already exhibit large negative magnetoresistance and unconventional Hall responses.\cite{bai201924,li201925,mazzola202327,li202632,su202607} Here we identify monolayer NbMnX$_2$ ($X$ = S, Se, Te) as an NC-FiM family in which the same canted spin state controls elastic stiffness, entropy, and electronic transport.

This design strategy differs from the usual search for a single optimized response. The Mn 3$d$ moments provide a large local-spin background, whereas the more delocalized Nb 4$d$ orbitals participate in both magnetism and the low-energy electronic bands. The Nb sublattice can therefore act simultaneously as a magnetic internal coordinate and an electronic control channel. Changing S, Se, or Te further tunes the orbital extent and SOC without changing the structural motif. The central question is whether these ingredients merely coexist or whether one microscopic magnetic state can connect them. By combining density functional theory (DFT), symmetry analysis, Wannier-function modeling, and atomistic Monte Carlo simulations, we show that the NC-FiM state provides a common microscopic link.

\section*{Crystal structure, magnetic ground state, and spin canting}

\begin{figure}[!ht]
  \centering
  \includegraphics[width=0.70\textwidth]{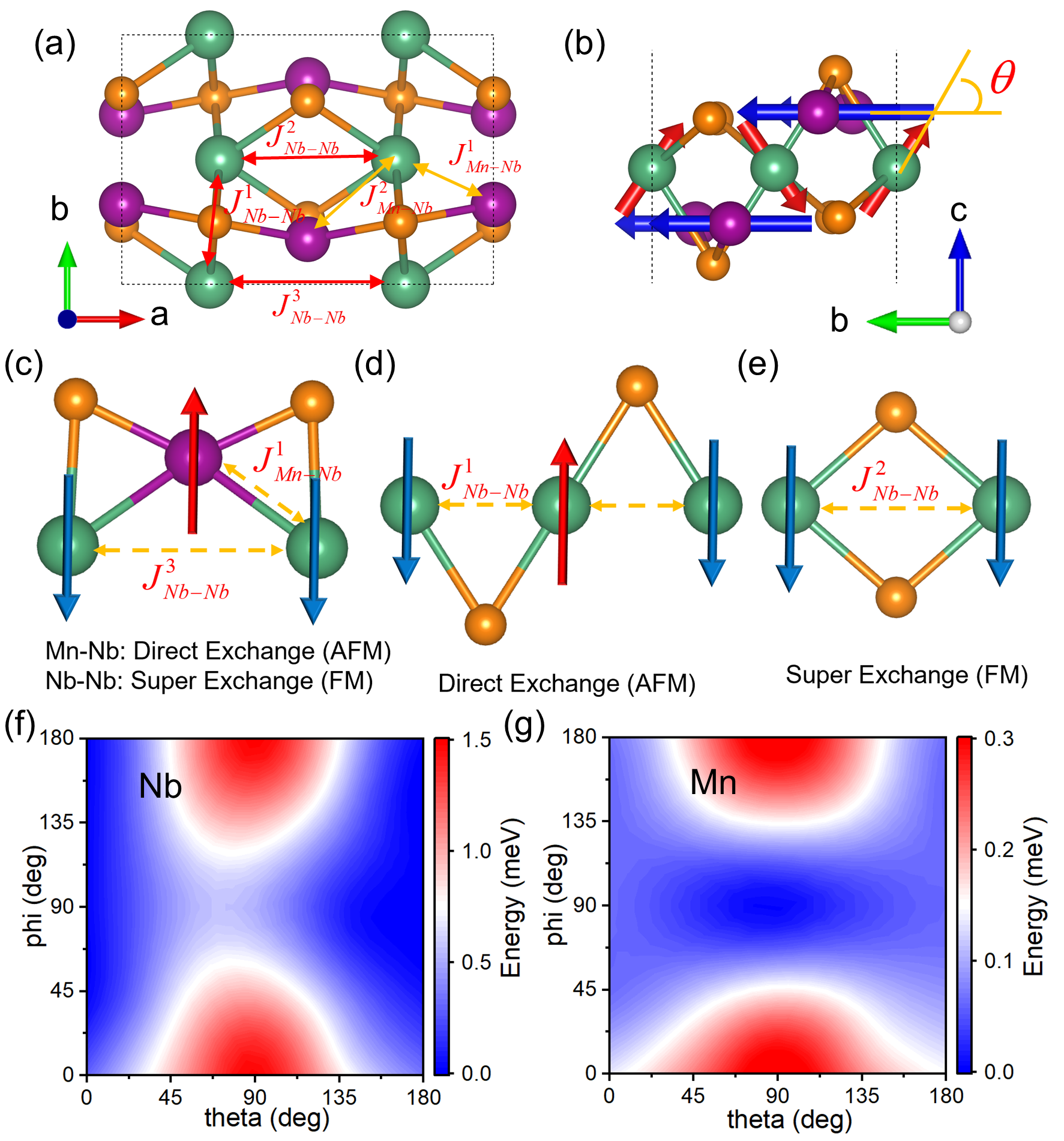}
  \caption{(a) Top view of NbMnX$_2$ with dominant exchange interactions marked by double-headed arrows. Nb, Mn, and X atoms are green, purple, and orange, respectively. (b) Side view and the ground-state spin configuration of NbMnTe$_2$; $\theta$ is the deviation of the Nb moment from the net magnetization axis. (c--e) Dominant exchange pathways. (f,g) Angle-dependent magnetocrystalline anisotropy energies of Nb and Mn in NbMnTe$_2$.}
  \label{fig:1}
\end{figure}

NbMnX$_2$ monolayers crystallize in the orthorhombic space group \textit{Pmna} [Fig.~\ref{fig:1}(a,b)]. Both transition-metal sublattices are magnetic: Mn carries more than 4 $\mu_{\mathrm B}$ and Nb more than 1 $\mu_{\mathrm B}$ (Table~\ref{tab:1}), unlike related NbM$''$X$_2$ compounds in which Nb is nearly nonmagnetic.\cite{bai201924,stepanova202429,wu202430,gao202531,ershadrad202633} Phonon spectra contain no imaginary modes, 300 K \textit{ab initio} molecular-dynamics trajectories retain their structures, and all formation energies are negative (Figs.~S1, S2, and Table~S1), supporting dynamical, thermal, and thermodynamic stability.

The energetic trend also supports chemical feasibility. The cohesive energies of NbMnS$_2$, NbMnSe$_2$, and NbMnTe$_2$ are 4.99, 4.61, and 4.26 eV/atom, respectively. The first two exceed that of experimentally synthesized monolayer NbFeTe$_2$ (4.53 eV/atom), while the Te compound remains comparable. Their formation energies range from $-2.18$ to $-1.82$ eV/atom and are more negative than reported values for experimentally realized CrI$_3$ and CrGeTe$_3$.\cite{huang201836} These comparisons do not by themselves establish a synthesis route, but together with the phonon and finite-temperature tests they show that the predicted magnetic states occur in well-defined metastable monolayers rather than mechanically unstable structures. The definitions and reference states used for these per-atom energies are given in Note 1 in the Supplemental Information (SI).

DFT identifies a noncollinear ferrimagnetic ground state throughout the family. The Mn moments align parallel, whereas the Nb moments are antiparallel to the net magnetization and cant away from it. NbMnS$_2$ and NbMnSe$_2$ prefer out-of-plane magnetization along $z$, with canting angles of 4.9$^\circ$ and 9.2$^\circ$, respectively; NbMnTe$_2$ prefers the in-plane $y$ direction and has a much larger 55.7$^\circ$ canting angle. The $x$ direction is the hard axis for all three compounds, and constraining the magnetization along $x$ restores a collinear ferrimagnetic state (Table~S2).

To identify the origin of these distinct canting angles, we first examine the competing exchange pathways. The first- and second-neighbor Mn--Nb interactions are strongly antiferromagnetic, with values below $-30$ and $-20$ meV, respectively, and Mn--Nb distances close to 3~\AA. The short bonds and spatially extended Nb 4$d$ orbitals favor direct exchange [Fig.~\ref{fig:1}(c)]. The nearest-neighbor Nb--Nb interaction is also antiferromagnetic, whereas the more distant $J_2^{\mathrm{Nb-Nb}}$ and $J_3^{\mathrm{Nb-Nb}}$ terms are ferromagnetic (Table~S3). The nearest-neighbor term is consistent with Goodenough--Kanamori--Anderson superexchange through a Nb--X--Nb bond angle near $90^\circ$, and its antiferromagnetic magnitude increases from S to Te as the chalcogen $p$ orbitals become more polarizable.\cite{goodenough195538} The longer-range terms can be understood as Mn-mediated interactions that tend to align neighboring Nb spins. Competition among these pathways creates a soft canting coordinate, even when isotropic exchange alone does not stabilize a nonzero angle.

We make this division of roles explicit with a minimal angle model that retains the two dominant Mn--Nb exchanges, the nearest-neighbor Nb--Nb exchange, and the first- and second-neighbor Mn--Nb Dzyaloshinskii--Moriya interaction (DMI) terms (Note 2 in the SI). With unit spin vectors and signed effective coefficients $A_1$, $A_2$, $B$, $D_1$, and $D_2$, the energy per formula unit is

\begin{equation}
E(\theta)=(A_1+A_2)\cos\theta-B\cos 2\theta-(D_1+D_2)\sin\theta-E_0 .
\end{equation}

The equilibrium canting follows from

\begin{equation}
-(A_1+A_2)\sin\theta+2B\sin 2\theta-(D_1+D_2)\cos\theta=0 .
\end{equation}

In the isotropic limit, a finite solution requires $\cos\theta=(A_1+A_2)/(4B)$, so frustration alone can cant the spins only when $|A_1+A_2|<|4B|$. Among the three compounds, this criterion is met only by NbMnTe$_2$. Numerical minimization gives 43.9$^\circ$ for NbMnTe$_2$ with isotropic exchange alone and 53.8$^\circ$ after DMI is included, in close agreement with the DFT value of 55.7$^\circ$ (Fig.~S3).

The DMI supplies the missing transverse torque in the S and Se compounds and enhances the exchange-driven canting in NbMnTe$_2$. It is symmetry allowed because the midpoint of a Mn--Nb bond is not an inversion center. The dominant $D_x$ component is perpendicular to the mainly $yz$ spin-rotation plane and therefore couples directly to the canting coordinate. Its magnitude grows from 1--2 meV in NbMnS$_2$ and NbMnSe$_2$ to 6--8 meV in NbMnTe$_2$ (Table~S4). For NbMnS$_2$, adding DMI to the otherwise collinear isotropic model yields 5.4$^\circ$, close to the DFT value of 4.9$^\circ$. For NbMnSe$_2$, the same reduced model yields 20.3$^\circ$, compared with 9.2$^\circ$ from DFT. The corresponding model energy curves and minima for both compounds are shown in Fig.~S4. A sensitivity check in which the nearest-neighbor Nb--Nb coefficient is changed from $-15.53$ to $-12.00$ meV shifts the model minimum to 10.4$^\circ$, close to the DFT result. This pronounced dependence shows that NbMnSe$_2$ lies near a frustrated canting instability, where the truncated model captures the mechanism but is sensitive to the exchange parameters retained. Thus, NbMnTe$_2$ is primarily frustration-canted and DMI-enhanced, whereas NbMnS$_2$ and NbMnSe$_2$ are primarily DMI-canted on exchange landscapes close to frustration.

Magnetic anisotropy selects the orientation of this canted state and stabilizes long-range 2D order. Because the in-plane axes are inequivalent, a uniaxial form is insufficient; for each magnetic species we use the orthorhombic single-ion Hamiltonian

\begin{equation}
H_{\rm ani}=-K_xS_x^2-K_yS_y^2-K_zS_z^2,
\end{equation}

with $K_x$, $K_y$, and $K_z$ extracted from atom-resolved SOC energies. The Nb contribution to the magnetocrystalline anisotropy exceeds 1 meV in NbMnTe$_2$, substantially larger than the Mn contribution [Fig.~\ref{fig:1}(f,g)], consistent with the extended Nb 4$d$ orbitals and their stronger SOC response; the same site dependence occurs in NbMnS$_2$ and NbMnSe$_2$ (Fig.~S5). The global hard direction is $x$, whereas the relative $y$ and $z$ energies select the compound-dependent easy axis. Using the calculated exchange, DMI, and orthorhombic anisotropy parameters in the classical spin Hamiltonian gives Curie temperatures of 292, 313, and 229 K for NbMnS$_2$, NbMnSe$_2$, and NbMnTe$_2$, respectively (Fig.~S6). These results indicate that isotropic exchange sets the canting stiffness, DMI supplies the transverse torque, and anisotropy confines the spin plane and supports finite-temperature order.

\begin{table}[htbp]
  \caption{Lattice parameters ($a$, $b$; \AA), on-site magnetic moments ($M_{\mathrm{Nb}}$, $M_{\mathrm{Mn}}$; $\mu_{\mathrm B}$), orthorhombic anisotropy parameters ($K_x$, $K_y$, $K_z$; meV), and Curie temperatures ($T_{\mathrm C}$; K) of NbMnX$_2$ monolayers.}
  \label{tab:1}
  \centering
  \resizebox{\textwidth}{!}{%
  \begin{tabular}{cccccccccccc}
    \toprule
    \multirow{2}{*}{System} & \multirow{2}{*}{$a$} & \multirow{2}{*}{$b$} & \multirow{2}{*}{$M_{\mathrm{Nb}}$} & \multirow{2}{*}{$M_{\mathrm{Mn}}$} & \multicolumn{3}{c}{Nb} & \multicolumn{3}{c}{Mn} & \multirow{2}{*}{$T_{\mathrm C}$} \\
    \cmidrule(lr){6-8} \cmidrule(lr){9-11}
    & & & & & $K_x$ & $K_y$ & $K_z$ & $K_x$ & $K_y$ & $K_z$ & \\
    \midrule
    NbMnS$_2$  & 8.63 & 5.89 & 1.35 & 4.37 & 0 & 0.25 & 1.06 & 0 & 0.06 & 0.08 & 292 \\
    NbMnSe$_2$ & 8.92 & 5.98 & 1.41 & 4.38 & 0 & 0.25 & 1.07 & 0 & 0.17 & 0.14 & 313 \\
    NbMnTe$_2$ & 9.27 & 6.16 & 1.35 & 4.43 & 0 & 0.97 & 1.53 & 0 & 0.31 & 0.25 & 229 \\
    \bottomrule
  \end{tabular}}
\end{table}

\section*{Magnetoelastic coupling and mechanical control}

The soft canting coordinate identified above provides a natural route for coupling magnetism to mechanics. As an initial signature, rotating NbMnTe$_2$ from its easy $y$ axis to the hard $x$ direction contracts $a$ by 0.58\% and expands $b$ by 0.46\%. This reorientation-induced deformation signals strong spin--lattice coupling, but it does not by itself distinguish two mechanically different effects: a shift of the equilibrium lattice and a change in the resistance to further deformation. To separate them, we write the strain-dependent energy of an orthorhombic sheet as

\begin{equation}
E_{\rm el}+E_{\rm me}=\tfrac12C_{11}\epsilon_{xx}^{2}+\tfrac12C_{22}\epsilon_{yy}^{2}+C_{12}\epsilon_{xx}\epsilon_{yy}+B_1\epsilon_{xx}+B_2\epsilon_{yy}+C_1\epsilon_{xx}^{2}+C_2\epsilon_{yy}^{2} .
\end{equation}

Here $C_{ij}$ are the reference elastic constants, whereas $B_i$ and $C_i$ are the linear- and quadratic-in-strain magnetoelastic coefficients, respectively. The $B_i$ terms displace the energy minimum and generate spontaneous magnetostriction; the $C_i$ terms change the curvature and hence renormalize the elastic stiffness. Minimizing the energy gives the magnetostriction coefficients $\lambda_x$ and $\lambda_y$, with the full derivation provided in Note 3 of the SI.

\begin{figure}[!ht]
  \centering
  \includegraphics[width=0.70\textwidth]{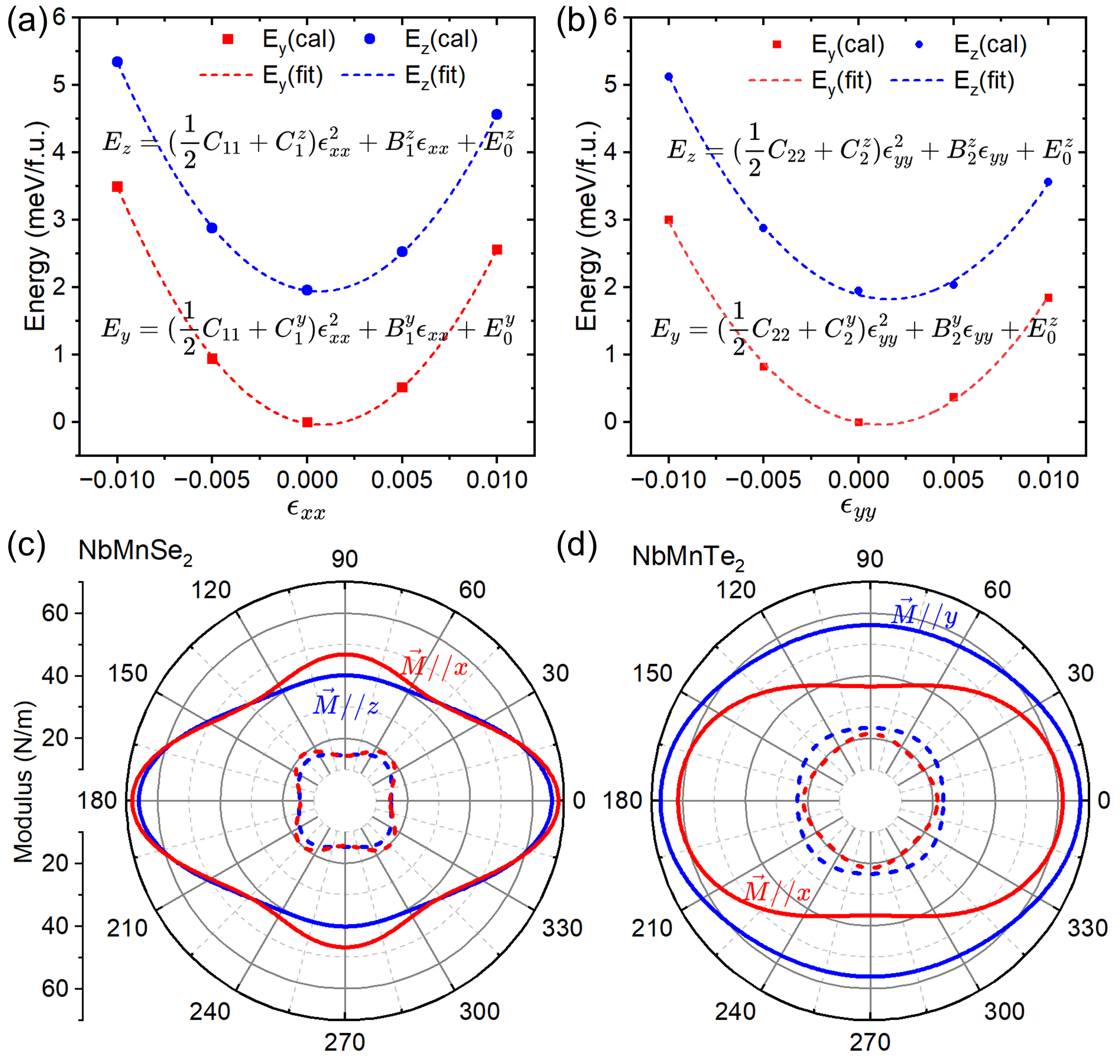}
  \caption{(a,b) Calculated energies (symbols) and fitted strain--energy curves (dashed lines) of NbMnTe$_2$ under uniaxial strain along (a) $x$ and (b) $y$. (c,d) Angular dependence of the in-plane Young's modulus $E(\theta)$ (solid) and shear modulus $G(\theta)$ (dashed) of (c) NbMnSe$_2$ and (d) NbMnTe$_2$. Red curves denote magnetization along $x$; blue curves denote magnetization along $z$ in (c) and along $y$ in (d).}
  \label{fig:2}
\end{figure}

We tested this framework by fitting magnetization-dependent DFT strain--energy curves. The model reproduces the NbMnTe$_2$ data in Fig.~\ref{fig:2}(a,b), as well as those of NbMnS$_2$ and NbMnSe$_2$ in Fig.~S7; the fitted coefficients are summarized in Table~S5. The linear response is strongly enhanced in NbMnTe$_2$, for which $B_1=8.96$ and $B_2=-25.2$ mJ m$^{-2}$, compared with magnitudes below 1 mJ m$^{-2}$ in NbMnS$_2$ and NbMnSe$_2$. The corresponding NbMnTe$_2$ magnetostriction coefficients are $\lambda_x=-235$ and $\lambda_y=484$ ppm. The latter exceeds the typical 200--400 ppm range of bulk Fe--Ga alloys.\cite{cao200940} The same fits yield sizable quadratic coefficients throughout the family, reaching $C_1=-511$ and $C_2=-148$ mJ m$^{-2}$ in NbMnTe$_2$ (Table~S5).

The large curvature response can be understood by treating the Nb canting angle $\theta$ as a strain-relaxable internal coordinate. Strain perturbs the Mn--Nb and Nb--Nb exchange interactions, DMI, and magnetic anisotropy, allowing $\theta$ to relax and feed back into the elastic energy. More generally, minimizing $E_{\rm mag}(\epsilon,\theta)$ with respect to $\theta$ gives

\begin{equation}
C_{\rm eff}=C_0+\frac{\partial^2E_{\rm mag}}{\partial\epsilon^2}-\frac{\left(\partial^2E_{\rm mag}/\partial\epsilon\partial\theta\right)^2}{\partial^2E_{\rm mag}/\partial\theta^2} .
\end{equation}

The final, negative term is the magnetic analogue of a relaxed-ion correction. Its denominator is the canting stiffness, so the elastic response can be enhanced when the internal spin mode softens. This behavior is consistent with a relaxed canting coordinate whose stiffness is modified by exchange and relativistic interactions (Note 3 in the SI).\cite{wu200541}

Direct elastic calculations verify the predicted stiffness control (Table~S5). NbMnS$_2$ changes only weakly upon reorientation. In NbMnSe$_2$, rotating the magnetization from $z$ to $x$ increases $C_{22}$ from 49.3 to 55.2 N m$^{-1}$. In NbMnTe$_2$, rotation from $y$ to $x$ reduces $C_{11}$ from 71.1 to 64.0 N m$^{-1}$ and $C_{22}$ from 59.5 to 38.1 N m$^{-1}$, a 36\% reduction in $C_{22}$. The angular Young's and shear moduli in Fig.~\ref{fig:2}(c,d) further show that the response is concentrated near the $y$ direction in NbMnSe$_2$ but extends over a broad angular range in NbMnTe$_2$. The family trend is therefore consistent with a cooperative role of SOC and a soft noncollinear coordinate, rather than with SOC strength or the equilibrium canting angle alone.

The response magnitude and the ease of magnetic control favor different compounds. NbMnSe$_2$ has the smaller reorientation energy, 2.40 meV per formula unit (f.u.), and is therefore the more accessible candidate for direct field-controlled stiffness modulation. NbMnTe$_2$ offers the larger mechanical response, but its 24.1 meV/f.u. barrier may require a magnetic field combined with strain bias from a flexible or piezoelectric substrate. Magnetization-dependent stiffness can be tested in suspended monolayer drums by resonance-frequency measurements or nanoindentation, and in supported samples by Brillouin or Raman spectroscopy or surface-acoustic-wave delay lines.\cite{lemme202042,mai202143,chen202344} Kerr rotation, X-ray magnetic circular dichroism, or nitrogen-vacancy magnetometry can monitor the magnetic state simultaneously.\cite{gong201745}

\section*{Spin--lattice enhancement of the magnetocaloric response}

The preceding elastic response establishes that magnetic order and lattice energetics are strongly coupled. We therefore ask whether this feedback also contributes to the entropy change near the magnetic transition. NbMnS$_2$ and NbMnSe$_2$ are the most relevant members for this purpose because their Curie temperatures, 292 and 313 K, lie near room temperature, whereas that of NbMnTe$_2$ is 229 K (Table~\ref{tab:1} and Fig.~S6). Their finite ferrimagnetic moments of approximately 3 $\mu_{\mathrm B}$/f.u. allow an applied field to modify the magnetic order parameter directly. We obtained $M(H,T)$ from Monte Carlo simulations for fields along the hard $x$ and easy $z$ directions and first evaluated the magnetic entropy change through the Maxwell relation

\begin{equation}
\Delta S_M(H,T)=\mu_0\int_0^H\left(\frac{\partial M}{\partial T}\right)_H\,\mathrm dH .
\end{equation}

\begin{figure}[!ht]
  \centering
  \includegraphics[width=0.70\textwidth]{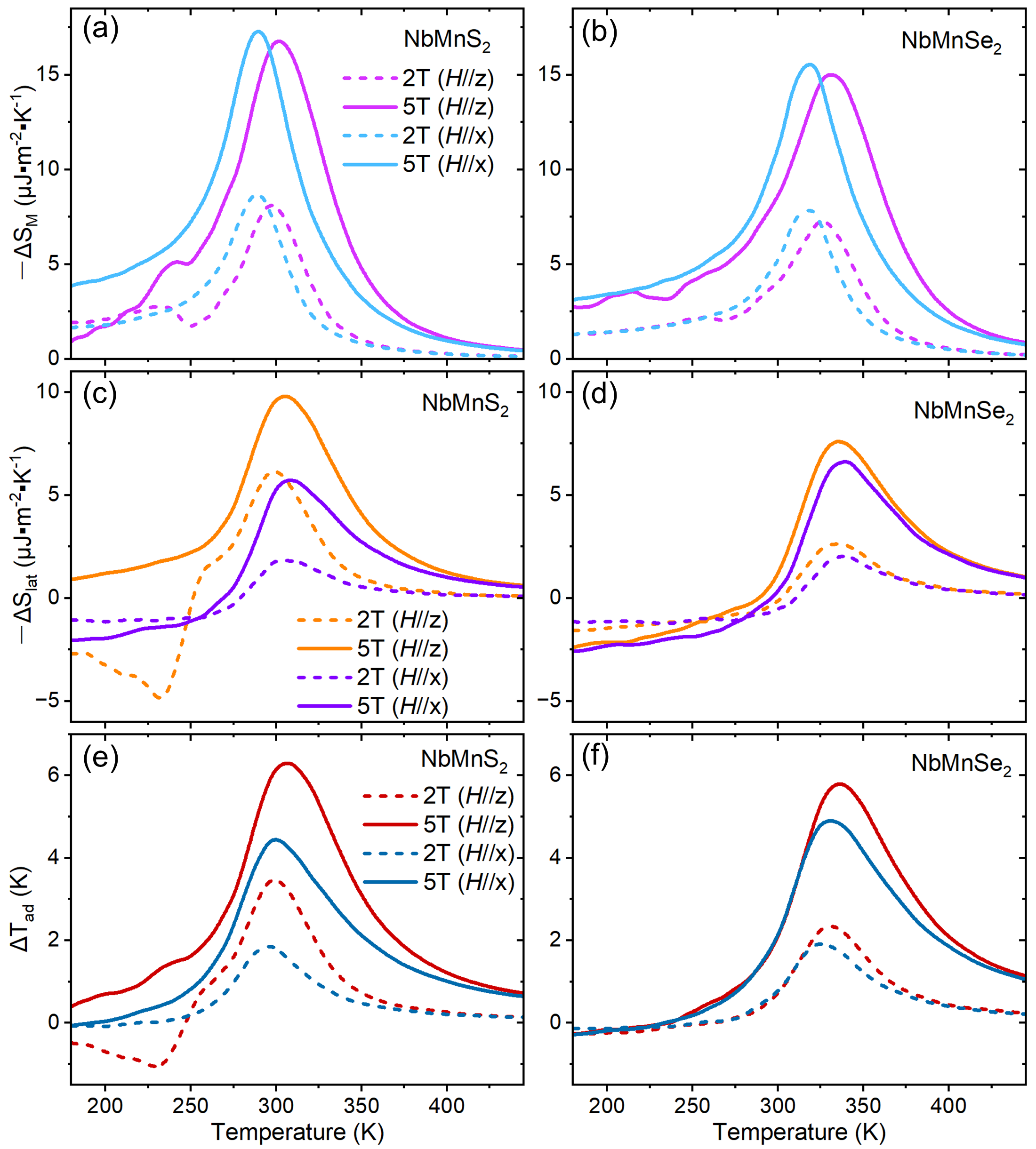}
  \caption{(a,b) Isothermal magnetic entropy change, (c,d) lattice entropy change, and (e,f) adiabatic temperature change of NbMnS$_2$ and NbMnSe$_2$. Results are shown for fields of 2 and 5 T applied along $z$ or $x$.}
  \label{fig:3}
\end{figure}

Under 5 T, the peak cooling magnitudes $-\Delta S_M$ are 17.3 and 16.8 $\mu$J m$^{-2}$ K$^{-1}$ for NbMnS$_2$ with $H\parallel x$ and $H\parallel z$, respectively, and 15.5 and 15.0 $\mu$J m$^{-2}$ K$^{-1}$ for NbMnSe$_2$ [Fig.~\ref{fig:3}(a,b)]. The corresponding peak temperatures are 289 and 301 K for NbMnS$_2$ and 319 and 330 K for NbMnSe$_2$. Thus, changing the field direction alters the peak magnitude only modestly but shifts its temperature by more than 10 K. This behavior follows from the distinct $M(H,T)$ profiles produced by hard- and easy-axis fields: through the Maxwell relation, their different temperature derivatives move the entropy maximum. Field orientation can therefore select the operating-temperature window without changing the composition. These peak values exceed the approximately 13 $\mu$J m$^{-2}$ K$^{-1}$ reported for CrI$_3$ under 5 T and occur close to room temperature.\cite{he202347}

A magnetic-only description, however, omits the lattice response established in the preceding section. First-principles phonon calculations show that the lattice entropy of the paramagnetic reference state exceeds that of the fully ferrimagnetic state over the temperature range considered (Fig.~S8). Loss of magnetic order can therefore increase the phonon entropy, providing a second entropy channel near the transition. To capture this lowest-order spin--lattice feedback, we interpolate between the two phonon endpoints using the simulated magnetization as the order parameter (Note 1 in the SI):

\begin{equation}
S_{\rm lat}(H,T)=\frac{M(H,T)}{M_0}S_{\rm FiM}(T)+\left[1-\frac{M(H,T)}{M_0}\right]S_{\rm PM}(T),
\end{equation}

where $S_{\rm FiM}$ and $S_{\rm PM}$ are obtained from first-principles phonons and $M_0$ is the ground-state moment. The field-induced lattice entropy change is then $\Delta S_{\rm lat}(H,T)=S_{\rm lat}(H,T)-S_{\rm lat}(0,T)$. This endpoint interpolation does not replace an explicit finite-temperature spin--phonon calculation, but it isolates the leading thermodynamic consequence of the calculated entropy difference between the ordered and disordered magnetic states.

The resulting lattice contribution changes sign below the transition but reinforces the magnetic entropy change in the critical region where the cooling response is largest [Fig.~\ref{fig:3}(c,d)]. At the temperatures of the 5 T magnetic-entropy maxima, it accounts for 14.0\% and 36.5\% of the total entropy change in NbMnS$_2$ for $H\parallel x$ and $H\parallel z$, respectively, and 20.2\% and 32.9\% in NbMnSe$_2$. Within the interpolation, this directional contrast follows from the different field-dependent order parameters entering $S_{\rm lat}(H,T)$. The larger fractions for $H\parallel z$ show that the lattice channel can substantially amplify the magnetocaloric response even when the magnetic entropy peaks themselves have similar magnitudes.

Finally, we combined the two entropy channels using the total heat capacity $C_p=C_{\rm mag}+C_{\rm lat}$ (Fig.~S9); the full expressions and sign convention are given in Note 1 of the SI. The resulting peak adiabatic temperature changes are 4.44 K ($H\parallel x$) and 6.29 K ($H\parallel z$) for NbMnS$_2$, and 4.90 and 5.79 K, respectively, for NbMnSe$_2$ [Fig.~\ref{fig:3}(e,f)]. The component-resolved curves show that the lattice channel contributes about half of the maximum out-of-plane temperature change (Fig.~S10). Even at 2 T, NbMnS$_2$ retains a 3.46 K response near 298 K for $H\parallel z$, whereas the response maximum of NbMnSe$_2$ lies above room temperature. These values describe the intrinsic response of a free monolayer; substrate heat capacity, thermal boundary resistance, and field-cycling losses will reduce the temperature swing of a device. Nevertheless, the sizable lattice contribution identifies a route to enhanced magnetocaloric performance that is absent from a magnetic-only model.

\section*{Magnetization-controlled metal--insulator transition and transport}

Having established the mechanical and thermal consequences of the canted magnetic state, we next ask whether magnetization rotation can also control the electronic phase. The family comparison in Table~S2 identifies NbMnSe$_2$ as the most favorable case. NbMnS$_2$ remains semiconducting, with gaps of 158 and 134 meV for $\hat M\parallel z$ and $\hat M\parallel x$, respectively. NbMnTe$_2$ switches from a semiconductor with a 243 meV gap for its easy-axis state $\hat M\parallel y$ to a metal for $\hat M\parallel x$, but the reorientation costs 24.1 meV/f.u. (Fig.~S11). NbMnSe$_2$ undergoes the same semiconductor-to-metal transition between $\hat M\parallel z$ and $\hat M\parallel x$ at a much smaller energy cost of 2.40 meV/f.u. It therefore combines a clear electronic contrast with a comparatively accessible magnetic reorientation.

\begin{figure}[!ht]
  \centering
  \includegraphics[width=0.70\textwidth]{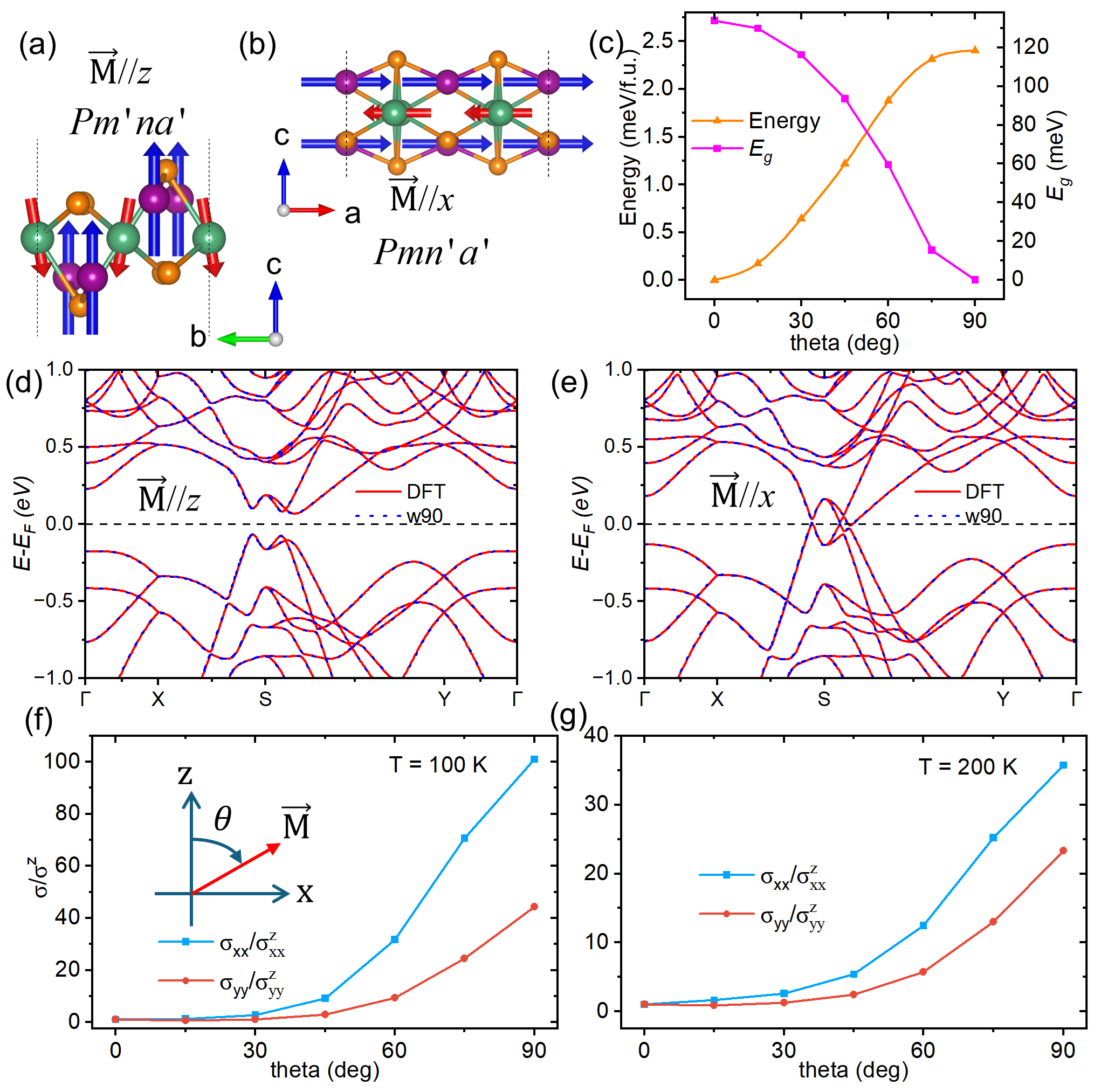}
  \caption{(a,b) NbMnSe$_2$ spin configurations for magnetization along (a) $z$ and (b) $x$. (c) Relative energy and band gap during magnetization rotation from $z$ to $x$. Here the plotted angle denotes the rotation of the net magnetization from $z$ and is distinct from the Nb spin-canting angle in Fig.~\ref{fig:1}. (d,e) DFT bands and Wannier fits (dotted) for magnetization along (d) $z$ and (e) $x$. (f,g) Conductivities normalized to their $\hat M\parallel z$ values during the same rotation at (f) 100 and (g) 200 K.}
  \label{fig:4}
\end{figure}

The two endpoints of this reorientation are shown in Fig.~\ref{fig:4}(a,b). The easy-axis $\hat M\parallel z$ state is noncollinear, with a Nb canting angle of 9.2$^\circ$, whereas the hard-axis $\hat M\parallel x$ state is a collinear ferrimagnet. We denote the angle through which the net magnetization rotates from $z$ toward $x$ by $\theta_M$ (labeled $\theta$ in Fig.~\ref{fig:4}), with $\theta_M=0^\circ$ and $\theta_M=90^\circ$ corresponding to these two states, respectively. Along this path, the energy rises smoothly to 2.40 meV/f.u. while the approximately 0.13 eV band gap closes continuously [Fig.~\ref{fig:4}(c)]. Combining this energy difference with the net moment of approximately 3 $\mu_{\mathrm B}$/f.u. gives an ideal coherent-rotation field scale of 13.8 T. This estimate characterizes the anisotropy-energy scale rather than a coercive field, which will depend on the actual switching pathway and domain structure.

The metal--insulator transition originates from a change in magnetic symmetry acting on the Nb-derived low-energy states. Orbital projections show that Nb 4$d$ states dominate near the Fermi level, whereas the Mn 3$d$ states lie farther away and change little during rotation (Fig.~S12). For $\hat M\parallel x$, the magnetic space group Pmn$^{\prime}$a$^{\prime}$ retains the unitary mirror $m_x$. The two bands at the Fermi-level crossing have opposite mirror eigenvalues and therefore cannot hybridize. Rotating the magnetization toward $z$ changes the magnetic space group to Pm$^{\prime}$na$^{\prime}$ and removes this protection, allowing SOC to hybridize the bands and open a gap. Magnetization rotation thus changes the allowed band coupling rather than merely shifting otherwise unchanged bands.

This symmetry mechanism can be summarized by a two-band $\mathbf{k}\cdot\mathbf{p}$ Hamiltonian near the crossing,

\begin{equation}
H(\mathbf q)=v_xq_x\sigma_x+v_yq_y\sigma_y+m(\mathbf M)\sigma_z,
\end{equation}

where $\mathbf q$ is measured from the crossing, $v_x$ and $v_y$ are the anisotropic band velocities, and the Pauli matrices act in the two-band subspace. The mirror symmetry $m_x$ requires $m(M_x,0,0)=0$, whereas breaking it permits $m(0,0,M_z)=\lambda M_z$ to leading order. The resulting gap, $E_g=2|\lambda M_z|$, therefore decreases continuously as $M_z$ vanishes during rotation toward $x$. For an approximately fixed moment, $M_z=M\cos\theta_M$, accounting for the continuous trend in Fig.~\ref{fig:4}(c). The full model and symmetry constraints are given in Note 4 of the SI.

Wannier interpolation accurately reproduces the DFT bands in both endpoint states [Fig.~\ref{fig:4}(d,e)], providing a basis for constant-relaxation-time Boltzmann transport calculations (Note 1 in the SI). At 100 and 200 K, both below the 313 K Curie temperature, $\sigma_{xx}$ increases monotonically as the gap closes and reaches 101 and 35.7 times its $\hat M\parallel z$ value, respectively [Fig.~\ref{fig:4}(f,g)]. The $y$-direction conductivity follows a nonmonotonic path: at $\theta_M=15^\circ$, $\sigma_{yy}$ falls to 66\% and 92\% of its initial value at 100 and 200 K, respectively, before reaching 44.3 and 23.3 times its initial value in the metallic $\hat M\parallel x$ state. This difference between $\sigma_{xx}$ and $\sigma_{yy}$ reflects the combined evolution of carrier availability and anisotropic band velocities in the orthorhombic lattice.

To quantify the orientation-induced conductivity contrast, we define the conductivity-based AMR ratio as\cite{rao202648}

\begin{equation}
\mathrm{AMR}_{ii}=\frac{|\sigma_{ii}^{x}-\sigma_{ii}^{z}|}{\sigma_{ii}^{z}}\times100\% .
\end{equation}

This definition gives $\mathrm{AMR}_{xx}=10{,}000\%$ and 3470\% at 100 and 200 K, respectively, and $\mathrm{AMR}_{yy}=4330\%$ and 2230\% at the same temperatures. These conductivity contrasts are far larger than the typical AMR scales of conventional ferromagnets, although numerical comparisons depend on whether conductivity or resistivity is used and on the chosen normalization.\cite{ritzinger202322} Within this model, rotation changes the electronic phase and carrier availability, rather than only modifying spin-dependent scattering within a fixed metallic state. Experimentally, strain bias, exchange bias, or coupling to a proximate magnetic layer could reduce the field needed to follow the coherent rotation path. Simultaneous electrical and Kerr or magnetic measurements could test the predicted continuous conductivity evolution and distinguish it from domain-wall motion or irreversible heating. The constant-relaxation-time results isolate the band-structure contribution; scattering and domains will determine the response magnitude in an actual device.

\section*{Conclusion}

In summary, first-principles calculations, symmetry analysis, atomistic spin simulations, and Wannier transport modeling identify NbMnX$_2$ monolayers as noncollinear ferrimagnets with Curie temperatures of 229--313 K. The absence of imaginary phonons, intact 300 K molecular-dynamics trajectories, and negative formation energies support their dynamical, thermal, and energetic stability, respectively; both Mn and Nb carry substantial moments. The minimal spin model separates the microscopic roles in canting: NbMnTe$_2$ is canted primarily by frustrated isotropic exchange and further tilted by DMI, whereas NbMnS$_2$ and NbMnSe$_2$ require symmetry-allowed DMI acting on exchange landscapes close to frustration. Orthorhombic magnetic anisotropy, with a strong contribution from the extended Nb 4$d$ states, selects the orientation of the canted state and supports finite-temperature order.

This field-responsive noncollinear order provides a common basis for three functional responses through distinct microscopic channels. First, magnetization rotation shifts the equilibrium strain and allows the canting angle to relax, producing a 36\% reduction in $C_{22}$ in NbMnTe$_2$. Second, coupling the field-dependent magnetic order to the difference between ferrimagnetic and paramagnetic phonon entropies yields a substantial lattice contribution and a peak adiabatic temperature change of 6.29 K near room temperature. Third, rotating NbMnSe$_2$ from $z$ to $x$ restores the $m_x$-protected band crossing and continuously closes an approximately 0.13 eV gap, increasing $\sigma_{xx}$ by a factor of 101 and producing a conductivity-based AMR ratio up to 10,000\% at 100 K. Together, these calculations suggest that a field-responsive noncollinear magnetic state can couple separately to strain, phonons, and low-energy electronic bands. Vector fields, strain bias, magnetic probes, nanomechanical measurements, and electrical transport provide complementary tests of these predictions.

\section*{Computational methods}

We performed spin-polarized DFT calculations using the PAW method in VASP with the PBE functional, a 500 eV plane-wave cutoff, a $9\times13\times1$ $k$ mesh, and Dudarev Hubbard parameters $U_{\mathrm{eff}}=5.25$ eV for Mn and 1.13 eV for Nb, obtained by linear response (Fig.~S13).\cite{kresse199949,kresse199650,perdew199651,dudarev199852,cococcioni200553} Structures were relaxed without symmetry constraints until all forces were below 0.01 eV~\AA$^{-1}$ and the total-energy change was below $10^{-5}$ eV; periodic images were separated by 15~\AA of vacuum. Finite-displacement phonons were calculated with Phonopy in $2\times2\times1$ supercells, from which the lattice entropy and heat capacity were obtained. Canonical-ensemble \textit{ab initio} molecular dynamics used the same supercell at 300 K for 10 ps with a 1 fs time step.\cite{togo202355,togo201056} Atom-resolved SOC matrix elements evaluated over magnetization orientations were used to determine the orthorhombic magnetic anisotropy. Magnetization-dependent strain--energy curves were fitted to the elastic and magnetoelastic expansion described in Note 3 of the SI.

Isotropic exchange and DMI parameters were extracted with the LKAG formalism in TB2J from SIESTA calculations based on the VASP-relaxed structures.\cite{he202157,jos200258,garca201859} Together with the orthorhombic anisotropy, these parameters define a classical spin Hamiltonian, which was simulated with VAMPIRE to obtain the Curie temperatures, field- and temperature-dependent magnetization, and magnetic heat capacity.\cite{evans201460} Magnetic entropy changes were evaluated from the Maxwell relation, and the lattice contribution was estimated by interpolating the phonon entropy between ferrimagnetic and paramagnetic reference states using the simulated magnetization as the order parameter.\cite{franco201846,martinhovieira202221} For NbMnSe$_2$, SOC-included bands were mapped onto maximally localized Wannier functions comprising Mn and Nb $d$ and Se $p$ orbitals. Conductivities were then calculated with WannierTools using the semiclassical Boltzmann equation in the constant-relaxation-time approximation and a $111\times171\times1$ mesh.\cite{pizzi202061,wu201862} The full spin Hamiltonian, Monte Carlo protocol, entropy treatment, and transport settings are given in Note 1 of the SI; the reduced canting and two-band models are detailed in Notes 2 and 4, respectively.

\section*{Acknowledgments}

This work was supported by the Fundamental Research Funds for the Central Universities (SWU-KT24007), National Natural Science Foundation of China (12274050, 12504207, and 12534012), and Guangdong Provincial Quantum Science Strategic Initiative (GDZX2401002). The authors acknowledge the high-performance computing platform of the School of Physical Science and Technology at Southwest University for providing computing resources.

\printbibliography

\end{document}